%% file: main.tex
\documentclass[conference]{IEEEtran}
\IEEEoverridecommandlockouts
\usepackage{cite}
\usepackage{amsmath,amssymb,amsfonts}
\usepackage{algorithmic}
\usepackage{booktabs}
\usepackage{graphicx}
\usepackage[acronym]{glossaries}
\usepackage{hyperref}
\usepackage[english]{babel}
\addto\extrasenglish{%
  \renewcommand{\figurename}{Fig.}%
}
\usepackage{standalone}
\usepackage{tikz}
\usetikzlibrary{decorations.pathmorphing}
\usepackage{textcomp}
\usepackage{xcolor}
\def\BibTeX{{\rm B\kern-.05em{\sc i\kern-.025em b}\kern-.08em
    T\kern-.1667em\lower.7ex\hbox{E}\kern-.125emX}}

\usepackage{fancyhdr}
\newacronym{awgn}{AWGN}{additive white Gaussian noise}
\newacronym{crlb}{CRLB}{Cramér-Rao lower bound}
\newacronym{csr}{CSR}{compressed sparse row}
\newacronym{fft}{FFT}{fast Fourier transform}
\newacronym{fim}{FIM}{Fisher information matrix}
\newacronym{gpu}{GPU}{Graphics Processing Unit}
\newacronym{jit}{JIT}{Just-In-Time}
\newacronym{ls}{LS}{least-squares}
\newacronym{nls}{NLS}{nonlinear least-squares}
\newacronym{mm}{MM}{matrix multiplication}
\newacronym{mse}{MSE}{mean squared error}
\newacronym{mle}{MLE}{maximum likelihood estimator}
\newacronym{pdf}{PDF}{probability density function}
\newacronym{snr}{SNR}{signal-to-noise ratio}

\begin{document}
\bstctlcite{IEEEexample:BSTcontrol}

\title{Fast Time-Domain MLE for\\ Period Estimation of Pulse Trains }

\author{\IEEEauthorblockN{
Sebastian~Schertler\IEEEauthorrefmark{1}\IEEEauthorrefmark{2}\IEEEauthorrefmark{3},
Daniel~Guger\IEEEauthorrefmark{1}\IEEEauthorrefmark{2}\IEEEauthorrefmark{3},
Stefan~Schuster\IEEEauthorrefmark{3},  
Stefan~Scheiblhofer\IEEEauthorrefmark{3},\\
Mario Huemer\IEEEauthorrefmark{1},
Alexander~Haberl\IEEEauthorrefmark{3},
Johann~Reisinger\IEEEauthorrefmark{3}, 
Oliver~Lang\IEEEauthorrefmark{1}\IEEEauthorrefmark{2}
}
\IEEEauthorblockA{\IEEEauthorrefmark{1}Institute of Signal Processing, Johannes Kepler University, 4040 Linz, Austria\\
\IEEEauthorrefmark{2}Christian Doppler Laboratory for Steel Industry Signal Processing and Machine Learning, Austria \\
\IEEEauthorrefmark{3}voestalpine Stahl GmbH, 4020 Linz, Austria\\
}}

\maketitle

\begin{abstract}
Parameter estimation of periodic pulse trains is a critical task in numerous automated sensing and diagnostic applications.
While estimation in the time domain provides superior accuracy in low signal-to-noise ratio environments, its high computational complexity frequently precludes its use in real-time systems.
This paper investigates algorithmic optimizations to reduce runtime by leveraging recent advancements in computing architectures.
Exploiting the inherent sparsity of the signal via sparse matrix multiplication kernels yields a substantial decrease in inference time.
Furthermore, by separating the dense matrix projections into sequential cross-correlation and sparse summation steps, we fundamentally reduce both runtime and memory complexity.
These optimizations drastically shrink the memory footprint, making time-domain estimation feasible for large datasets in real-time settings.
\end{abstract}

\color{red}
© 2026 IEEE.  Personal use of this material is permitted.  Permission from IEEE must be obtained for all other uses, in any current or future media, including reprinting/republishing this material for advertising or promotional purposes, creating new collective works, for resale or redistribution to servers or lists, or reuse of any copyrighted component of this work in other works.
\color{black}
\fancyfoot[C]{© 2026 IEEE}

\section{Introduction}
Periodic pulse parameter estimation is a fundamental task in applications ranging from industrial production processes \cite{RANDALL2011485} to time-of-arrival estimation in radar systems \cite[p.~247]{Richards2013-tl}.
In these applications, the acquired measurements frequently consist of short, periodic pulses separated by extended gaps consisting  of noise only.

Traditional approaches to periodic signal estimation often rely on frequency-domain techniques, such as the fundamental frequency (or pitch) estimator of multiharmonic signals \cite{Christensen.2009b}.
This estimator benefits from the computationally efficient \gls{fft} \cite{CooleyTukey} in the case of continuous sinusoidal signals., the case of short pulses results in wide bandwidths in the frequency domain, requiring a large model order. 
Additionally, as such signals may have a long period, a high degree of zero-padding may be needed to obtain estimates in the desired resolution. Knowledge of the pulse form can be included by weighting spectral amplitudes \cite{11226507}, improving the estimation performance in the low \gls{snr} region.

Based on a signal model in the time domain, we previously derived a \gls{mle} that asymptotically attains the \gls{crlb} \cite{Schertler.2025}.
A primary limitation of this estimator is the requirement of a computationally expensive grid search, limiting its applicability for long data records.

In this work, we investigate how to make this estimator computationally feasible by first optimizing the algorithmic complexity to reduce runtime and memory, and then focusing on the implementation itself.
Especially in recent years, propelled by the popularity of artificial intelligence and large language models, the hardware used to train and run these models has advanced rapidly.
Currently, GPUs are frequently used not only to render virtual scenes but also to run complex simulations, generate text, and perform other scientific workloads \cite{CHE20081370,10.1145/1654059.1654078}.
Our aim is to leverage recent advancements and optimize the periodic pulse estimator specifically for heavily parallelized hardware to ensure that the use of the estimator in real-time industrial processes becomes feasible.

This work is structured into four parts. First, we summarize the applied signal model and the \gls{mle}. Then we analyze the runtime and memory complexity of the baseline algorithm and the effects when sparsity and the matrix structure itself are taken into account. Various implementations are then tested to verify previous results and to detail their execution. Finally, we analyze the results and discuss the learned insights.
\section{Signal Models and Estimators}
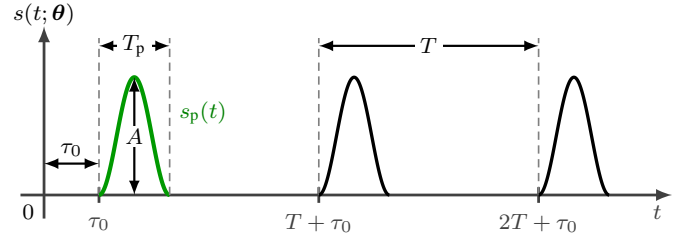
\begin{figure}[!t]
    \centering
    \resizebox{\columnwidth}{!}{\input{fig/pulses.tex}}
    \caption{Illustration of the signal model: periodically repeated pulses $s_\text{p}(t)$ multiplied by an unknown scaling factor $A$.}
    \label{fig:pulses}
\end{figure}
We assume a signal model consisting of the sum of $K$ periodically shifted, non-overlapping pulses $s_\text{p}(t)$ that are additionally scaled by $A$ \cite{Schertler.2025}:
\begin{align}
    s(t;\boldsymbol{\theta}) &= A\sum_{k=0}^{K-1} s_\text{p}(t-kT-\tau_0).
\end{align}
$\tau_0 \in [0;T)$ denotes the initial delay of the signal, with $T$ being the period.
$\boldsymbol{\theta} = [A\,\, T\,\, \tau_0]^T$ denotes the vector containing all unknown parameters.
The pulses $s_\text{p}(t)$ are defined to be zero outside $0 \leq t \leq T_\text{p}$ with $T_\text{p} \ll T$.
The resulting signal $s(t;\boldsymbol{\theta})$ is therefore zero between pulses.
$N$ samples with sampling interval $T_\text{s}$ are gathered into a vector $\mathbf{s}(\boldsymbol{\theta})$ and distorted by \gls{awgn}, which forms the measurement vector \mbox{$\mathbf{x} = \mathbf{s}(\boldsymbol{\theta}) + \mathbf{w}$}. \autoref{fig:pulses} visualizes the general structure of the signals.

\begin{figure}
    \centering
    \includegraphics[width=\linewidth]{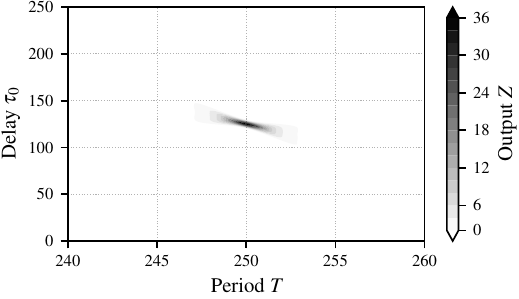}
    \caption{A cost function evaluated at various grid points. The true period $T$ is $250$ and the true delay $\tau_0$ is $125$. For large areas of the parameter space, the cost function is zero for the noiseless case. A fine grid is required as estimation is not feasible otherwise.}
    \label{fig:cost_func}
\end{figure}

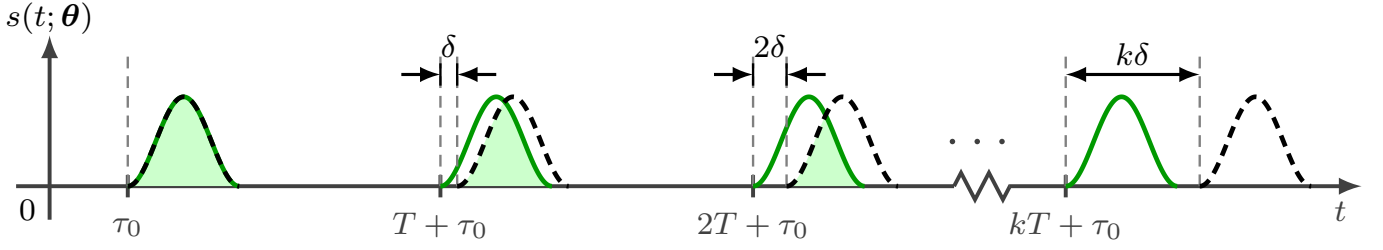
\begin{figure*}[!t]
    \centering
    \resizebox{\textwidth}{!}{\input{fig/drift.tex}}
    \caption{With an increasing number of pulses, a small difference between the assumed  $\tilde{T}$  and true period $T$ results in missed pulses.}
    \label{fig:drift}
\end{figure*}

For a signal in which the model is nonlinear in some parameters $\boldsymbol{\beta} = [T\, \tau_0]^T$ but linear in others, such as $\alpha = [A]$, the model can be reformulated into a separable least squares problem \cite[p.~223]{StevenM.Kay.1993}:
\begin{align}
    \mathbf{x}&=\underbrace{\mathbf{h(\boldsymbol{\beta})}}_{\mathbf{s}(\boldsymbol{\theta})|_{A=1}}\hspace{-2mm} \cdot\, \alpha + \mathbf{w}. \label{eq:meas_model}
\end{align}
As the model assumes \gls{awgn}, an estimator minimizing the \gls{ls} error is also the asymptotically efficient \gls{mle} \cite[p.~254]{StevenM.Kay.1993} under certain regularity conditions. The parameters that affect the signal linearly can be found by solving $\widehat{\alpha} = \left(\mathbf{h(\boldsymbol{\beta})}^T\mathbf{h(\boldsymbol{\beta})}\right)^{-1}\mathbf{h(\boldsymbol{\beta})}^T\mathbf{x},\label{eq:alpha_hat}$ which depends on $\boldsymbol{\beta}$. 
To estimate the remaining parameters (the period $T$ and delay $\tau_0$ in our case), one needs to maximize:
\begin{align}
    \boldsymbol{\widehat{\beta}} &= \arg\max_{\boldsymbol{\beta}} \underbrace{\mathbf{x}^T \mathbf{h(\boldsymbol{\beta})}\left(\mathbf{h(\boldsymbol{\beta})}^T\mathbf{h(\boldsymbol{\beta})}\right)^{-1}\mathbf{h(\boldsymbol{\beta})}^T\mathbf{x}}_{J(\boldsymbol{\beta})}. \label{eq:beta_hat}
\end{align}
In practice, it is not straightforward to implement this estimator efficiently.
As the pulses are short and most samples consist of noise only, the support in the parameter space is very limited.
Consequently, gradient-based optimization is ineffective, as the objective function yields a vanishing gradient of near zero across the majority of the parameter space.
\autoref{fig:cost_func} shows the cost function $J(\boldsymbol{\beta})$  evaluated on a grid of various periods $T$ and delays $\tau_0$, illustrating the small support region.
Additionally, for measurements with many pulses, a fine grid becomes increasingly important.
The term in \eqref{eq:beta_hat} effectively calculates the energy in the measurement signal explained by the model parameterized by $\boldsymbol{\beta}$.
When the true period $T$ and the assumed period $\tilde{T}$ do not match exactly ($\tilde{T}-T  = \delta$), the explained energy is reduced due to a lack of overlap, even for the correct delay $\tau_0$.
For a large number of pulses, this may lead to the loss of entire pulses (\autoref{fig:drift}).
This behavior can also be explained by looking at the \gls{fim} \cite{schertler2025estimatorsperformanceboundsshort}.
With $N_\text{p}=\lceil \frac{T_\text{p}}{T_\text{s}}\rceil$ as the number of non-zero samples for a single pulse, the diagonal entries for the period and delay are:
\begin{align}
    \left[ \mathbf{I}(\boldsymbol{\theta}) \right]_{T,T} &= A^2 \frac{2K^3-3K^2+K}{6} \cdot \sum_{n=0}^{N_\text{p}-1} \left(  \frac{\partial s_\text{p}(t) }{\partial t} \bigg|_{t=n T_\text{s}} \right)^2 \\
    \left[ \mathbf{I}(\boldsymbol{\theta}) \right]_{\tau_0, \tau_0}  &= A^2 K \cdot \sum_{n=0}^{N_\text{p}-1} \left(  \frac{\partial s_\text{p}(t) }{\partial t} \bigg|_{t=n T_\text{s}} \right)^2. 
\end{align}
The ratio between the elements grows quadratically with the number of pulses $K$:
\begin{align}
    \frac{\left[ \mathbf{I}(\boldsymbol{\theta}) \right]_{T,T}}{\left[ \mathbf{I}(\boldsymbol{\theta}) \right]_{\tau_0, \tau_0} } &= \frac{2K^2-3K+1}{6}.
\end{align}
The \gls{fim} effectively tells us how sensitive the joint \gls{pdf} is w.r.t. the parameters in question.
Thus, with an increasing number of pulses, the estimation accuracy for the period $T$ increases much faster, but it is also more sensitive with regard to a mismatch in parameters.
It is therefore necessary to increase the resolution along $T$, while for $\tau_0$, the grid density is less of an issue. 
\section{Algorithms and complexity}
In this section, we investigate how to reduce the runtime and memory complexity when calculating the cost function \eqref{eq:beta_hat}.
We apply the "Big $\mathcal{O}$" notation \cite{bachmann1921zahlentheorie,landau1909handbuch} as we are mainly interested in how the execution time and memory requirements scale with the number of parameter grid points $G$, $N$, $K$ and $N_\text{p}$.
\subsection{Base algorithm}
For the complexity of the baseline, we first derive $\mathbf{v}(\boldsymbol{\beta}) = \mathbf{h}(\boldsymbol{\beta})^T\mathbf{x}$ and $\mathbf{m}(\boldsymbol{\beta}) = \left( \mathbf{h}(\boldsymbol{\beta})^T\mathbf{h}(\boldsymbol{\beta})\right)^{-1}$, then $\mathbf{u}(\boldsymbol{\beta}) = \mathbf{m}(\boldsymbol{\beta})\mathbf{v}(\boldsymbol{\beta})$, and finally $J(\boldsymbol{\beta}) = \mathbf{v}^T\mathbf{u}$. 
This way, calculating and storing the massive $N\times N$ projection matrix $\mathbf{P}_\text{A}=\mathbf{h(\boldsymbol{\beta})}\left(\mathbf{h(\boldsymbol{\beta})}^T\mathbf{h(\boldsymbol{\beta})}\right)^{-1}\mathbf{h(\boldsymbol{\beta})}$ can be avoided, with the added benefit that one gets $\widehat{\alpha}=\mathbf{u}$  as an intermediate result.
\begin{table}[htbp]
\centering
\caption{Computational Complexity of Intermediate Results \\ (Baseline Algorithm)}
\label{tab:complexity_baseline_p1}
\begin{tabular}{lcc}
\toprule
Calculation & Runtime $\mathcal{O}(\cdot)$ & Memory $\mathcal{O}(\cdot)$ \\
\midrule
$\mathbf{h}(\boldsymbol{\beta})$ & $N$ & $N$ \\
$\mathbf{x}$ & $-$ & $N$ \\
$\mathbf{v}(\boldsymbol{\beta}) = \mathbf{x}^T\mathbf{h}(\boldsymbol{\beta})$ & $N$ & $1$ \\
$\mathbf{m}(\boldsymbol{\beta}) = \mathbf{h}(\boldsymbol{\beta})^T\mathbf{h}(\boldsymbol{\beta})$ & $N$ & $1$ \\
$\mathbf{m}^{-1}(\boldsymbol{\beta})$ & $1$ & $1$ \\
$\mathbf{u}(\boldsymbol{\beta}) = \mathbf{v}(\boldsymbol{\beta})\mathbf{m}^{-1}(\boldsymbol{\beta})$ & $1$ & $1$ \\
$J(\boldsymbol{\beta}) = \mathbf{u}(\boldsymbol{\beta})\mathbf{v}(\boldsymbol{\beta})^T$ & $1$ & $1$ \\
\midrule
Total per point & $N$ & $N$ \\
Total per $G$ grid points & $GN$ & $GN$ \\
\bottomrule
\end{tabular}
\end{table}
\subsection{Leveraging sparsity}
As the signal is sparse, many multiplications contain at least one zero, resulting in a large overhead.
Incorporating the structure in $\mathbf{h}(\boldsymbol{\beta})$ reduces the complexity of $\mathbf{h}(\boldsymbol{\beta})^T\mathbf{x}$ to $\mathcal{O}(K N_\text{p})$.
While straightforward, this modification heavily reduces the number of computations required.
Applying a sparse-matrix format like \gls{csr} \cite{1447941} results in a reduction of the runtime by a factor of $\frac{KN_\text{p}}{N}$ (the ratio of non-zero values to all samples).
For large data records, sparsity also makes it feasible to calculate $\mathbf{h}(\boldsymbol{\beta})$ and $(\mathbf{h}(\boldsymbol{\beta})^T\mathbf{h}(\boldsymbol{\beta}))^{-1}$ beforehand, resulting in a further decrease in runtime during inference with the trade-off of increased memory complexity $\mathbf{h}(\boldsymbol{\beta})^T=\mathcal{O}(GKN_\text{p})$.
\subsection{Correlation + Gather}
Although the sparse model strongly reduces the required number of calculations and also the memory, the matrix $\mathbf{h}(\boldsymbol{\beta})$ still consists of $\approx KG$ pulses $s_\text{p}(t)$ sampled at slightly different positions. 
Looking closer, the operation $\mathbf{h}(\boldsymbol{\beta})^T\mathbf{x}$ approximately correlates the measurement with the known pulse shape and sums the results periodically.
For multiple grid points, similar correlation operations may occur for different $\boldsymbol{\beta}$.
For instance, regardless of the period $T$, the first pulse will always be at the same location when $\tau_0$ is constant. Redundancies can be reduced by splitting $\mathbf{h}(\boldsymbol{\beta})^T\mathbf{x}$ into a correlation and a summation (or gathering) operation.
But as the period is not strictly bound to be an integer multiple of the sampling time, every pulse may be sampled at slightly different times, and hence, the correlation results differ from pulse to pulse. 
The possible shift is limited to $\varepsilon \in [-\frac{T_\text{s}}{2},\frac{T_\text{s}}{2}]$. Therefore, the pulse can be approximated using a Taylor series: 
\begin{align}
    s_\text{p}(t+\varepsilon) &=  s_\text{p}(t) + \varepsilon \frac{\partial s_\text{p}(t)}{\partial t} + \mathcal{O}(\varepsilon^2).
\end{align}
The approximation can be implemented efficiently by correlating the measurements with the pulse values and their derivatives sampled at $\varepsilon=0$. 
Subsequently, the gathering operation can be performed efficiently by multiplying the correlation result with a sparse matrix $\tilde{\mathbf{h}}(\boldsymbol{\beta}) \in \mathbb{R}^{N \times 2}$.
Compared to $\mathbf{h}(\boldsymbol{\beta})$, this structure decouples the discrete pulse locations from their continuous sub-sample shifts.
Specifically, the matrix is constructed as $\tilde{\mathbf{h}}(\boldsymbol{\beta}) = \begin{bmatrix} \mathbf{p} & \mathbf{e} \end{bmatrix}$, where the first column is a binary vector storing the nearest-sample pulse positions with ones and zeros. The second column $\mathbf{e} \in \mathbb{R}^{N \times 1}$ contains the corresponding sub-sample shifts $\varepsilon $ at those active positions and is zero elsewhere.

Note that the accuracy of the Taylor approximation strictly depends on the bandwidth of the pulse $s_\text{p}$.
This approximation is only valid if the maximum sub-sample shift $\varepsilon_\text{max}=\frac{T_\text{s}}{2}$ does not significantly change the phase of the highest frequency component. If a higher accuracy is desired one can simply use a higher order approximation with the added cost of runtime complexity.

By separating the correlation and gathering steps, one effectively decouples the impact of the number of measurements from the number of grid points (see \autoref{tab:complexity_gather_p1}) on the complexity.
In earlier algorithms, the runtime complexity depended on the product $GN$.
Now the runtime complexity scales linearly with the sum of these terms.
\begin{table}[htbp]
\centering
\caption{Computational Complexity of Intermediate Results \\ (Corr + Gather Algorithm)}
\label{tab:complexity_gather_p1}
\begin{tabular}{lcc}
\toprule
Calculation & Runtime $\mathcal{O}(\cdot)$ & Memory $\mathcal{O}(\cdot)$ \\
\midrule
$\mathbf{x}$ & $-$ & $N$ \\
$\mathbf{s}_p$ & $-$ & $N_\text{p}$ \\
$\tilde{\mathbf{h}}(\boldsymbol{\beta})$ (Sparse) & $K$ & $K$ \\
$\tilde{\mathbf{v}} = \mathbf{x} \star \mathbf{s}_p$ & $N N_\text{p}$ & $N$ \\
$\mathbf{m}(\boldsymbol{\beta}) = \mathbf{h}(\boldsymbol{\beta})^T\mathbf{h}(\boldsymbol{\beta})$ & $K$ & $1$ \\
$\mathbf{m}^{-1}(\boldsymbol{\beta})$ & $1$ & $1$ \\
$J(\boldsymbol{\beta}) = (\tilde{\mathbf{v}} \tilde{\mathbf{h}}(\boldsymbol{\beta}))^2 \mathbf{m}^{-1}(\boldsymbol{\beta})$ & $K$ & $1$ \\
\midrule
Total per point & $N N_\text{p} + K$ & $N + N_\text{p} + K$ \\
Total for $G$ points & $N N_\text{p} + G K$ & $N + N_\text{p} + G K$ \\
\bottomrule
\end{tabular}
\end{table}

An overview of all runtime and memory complexities can be found in \autoref{tab:algorithm_comparison}.
\begin{table}[t]
\centering
\caption{Architectural Comparison of Grid Evaluation Algorithms}
\label{tab:algorithm_comparison}
\begin{tabular}{lcc}
\toprule
Algorithm & Total Runtime $\mathcal{O}(\cdot)$ & Total Memory $\mathcal{O}(\cdot)$ \\
\midrule
Dense MM& $G N$ & $G N$ \\
Sparse MM& $G KN_\text{p}$ & $G KN_\text{p}$ \\
Correlation + Gather & $N N_\text{p} + G K$ & $N + N_\text{p} + G K$ \\
\bottomrule
\end{tabular}
\end{table}
\section{Implementation and Results}
For testing and benchmarking the algorithms detailed in the last section, we use the deep-learning-focused open-source library PyTorch \cite{paszke2019pytorchimperativestylehighperformance}.
Although several alternatives exist (like the libraries Numba, Jax, and Triton, or the programming language Julia), PyTorch was selected for several reasons. Not only is it heavily optimized for matrix multiplications and convolutions, but it also allows for using the same function on different computing devices from different vendors and includes benchmarking tools for accurately measuring execution time and profiling memory operations.

While the previous section focused on algorithmic complexity, practical performance is heavily dictated by hardware execution and system-level bottlenecks.
Over the past decade, the rapid advancement of machine learning has been intrinsically linked to the evolution of hardware accelerators.
Operations fundamental to deep learning, particularly large-scale matrix multiplications, are executed with significantly higher throughput on GPUs compared to CPUs due to highly parallelized architectures.

However, raw compute power is rarely the primary limiting factor. Performance is often bottlenecked by memory bandwidth. It is critical to minimize data transfers between system RAM and the GPU across the PCIe bus, as well as intermediate read/write operations to the GPU's own VRAM. 
To mitigate these overheads in computationally critical pipelines, one can use \gls{jit} compilation features like \verb|torch.compile|. Rather than routing individual operations to specific hardware components—which is already handled by underlying backend libraries—compilation analyzes the complete, step-by-step sequence of mathematical operations ahead of time. It identifies sequential operations and merges them into a single, unified set of GPU instructions. This consolidation prevents the hardware from writing intermediate results back to the main video memory (VRAM) between every step, drastically reducing memory access bottlenecks and eliminating the overhead of the CPU constantly scheduling new, individual tasks.
The parameters used for the benchmark can be found in \autoref{tab:benchmark_params}. 
\begin{table}[htbp]
\centering
\caption{Updated Benchmark Parameters}
\label{tab:benchmark_params}
\begin{tabular}{lll}
\toprule
\textbf{Parameter} & \textbf{Value} & \textbf{Description} \\ \midrule
$N$                & $20\,000$          & Number of samples \\
$T_\text{s}$       & $1.0$ s            & Sampling period \\
Pulse Type         & Hann Pulse         & Pulse function \\
Pulse Length       & $50$ s             & Length of the Hann pulse \\
$T$                & $1000$ s           & True period $T$ (nominal) \\
$\tilde{T}$        & $[975, 1025]$ s    & Grid range for $T$ \\
$T_\text{res}$     & $0.1$ s            & Resolution of $T$ \\
$\tau_{0,\text{res}}$ & $3$ s           & Resolution of $\tau_0$ \\
$N_\text{runs}$    & $1000$             & Number of runs\\
$B$                & $16$               & Global batch size \\
\bottomrule
\end{tabular}
\end{table}

For the baseline algorithms, we implemented two versions. The first calculates the matrix $\mathbf{h}(\boldsymbol{\beta})$ during inference when needed to save memory. All cost function values are calculated in parallel using vectorized mapping. Provided sufficient memory is available, precalculating $\mathbf{h(\boldsymbol{\beta})}$ and using batched \gls{mm} is recommended, as PyTorch can reduce computational overhead by employing the very efficient "cuBLAS" library \cite{cublas_web}. This also improves inference time when processing measurements because $\mathbf{h}(\boldsymbol{\beta})$ is constant and can be calculated once beforehand to reduce runtime during inference. 

\subsection{Runtime}
To ensure accurate execution time measurements, a warm-up iteration is performed for each function to account for memory allocation and JIT compilation overhead.
Memory measurements were performed in separate runs so that the runtime was not affected.
It is important to note that the following results not only reflect differences driven by the different implementations but are also affected by the hardware used during tests.
To provide a best-case scenario for both CPU and GPU based implementations, we run all tests on modern high-end hardware.
The server utilized was equipped with an AMD Threadripper PRO 9975WX with 32 cores and an NVIDIA RTX 6000 Ada with 48 GB of VRAM.

\begin{figure}[h]
    \centering
    \includegraphics[width=\linewidth]{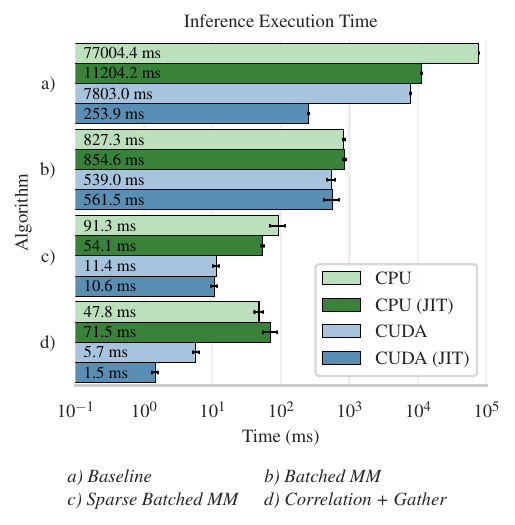}
    \caption{Inference runtime comparison of algorithms (standard and \gls{jit} versions) on a CPU and a GPU (CUDA).}
    \label{fig:runtime_inference}
\end{figure}

The results (\autoref{fig:runtime_inference}) show that any implementation using batched \gls{mm} (b) - d)) is faster than processing grid points separately when using a GPU (a)).
Additionally reducing the number of multiplications by employing sparse matrices (in c) and d)) strongly reduces the execution time.
While it is recommended to use sparse and batched \gls{mm} when using only a CPU, the performance of the compiled correlation + gather implementation is more than five times faster when a GPU is available.
As convolutions have been the backbone for many neural networks, improvements in GPU hardware and algorithmic domains are strongly visible here.

\subsection{Memory}
\begin{figure}
    \centering
    \includegraphics[width=\linewidth]{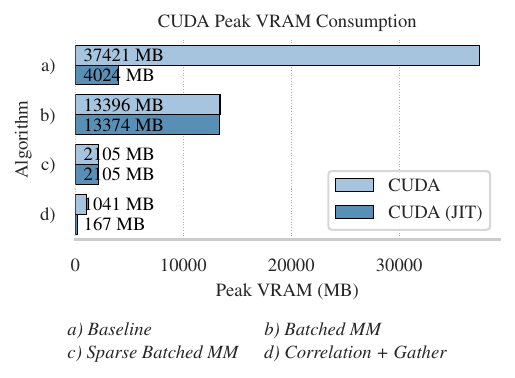}
    \caption{VRAM consumption for all CUDA implementations.}
    \label{fig:memory_inference}
    \vspace{-4mm}
\end{figure}

We evaluate the VRAM consumption of the GPU implementations (\autoref{fig:memory_inference}), as it provides a highly isolated metric that is largely unaffected by background system processes.
The uncompiled baseline requires over $37$\,GB to store massive intermediate projection matrices; however, \gls{jit} compilation fuses these steps, reducing memory by $89\%$. 
While dense batched \gls{mm} uses $13.4$ GB to store zeros and remains unaffected by compilation, applying sparse matrices strongly reduces the required memory. 
Ultimately, the correlation and gather approach consumes only $1$\,GB, and subsequent \gls{jit} compilation further optimizes its sequential operations to achieve a minimal footprint of just $167$\,MB.

\subsection{Estimation Performance}
To verify that the discussed algorithms do not degrade estimation performance, we simulated the \gls{mse} performance across various noise levels. We also included the multiharmonic pitch estimator \cite{Christensen.2009b} and a variant which additionally weights spectral amplitudes \cite{11226507}, allowing to include the pulse form itself.

\begin{figure}
    \centering
    \includegraphics[width=\linewidth]{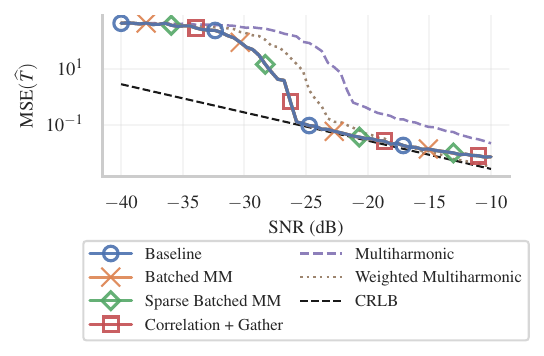}
    \caption{\gls{mse} performance over different \gls{snr}s. All implementations give very similar results. Methods not using pulse shortness performs worse in low \gls{snr}, but also attain the \gls{crlb} for high \gls{snr} regions.}
    \label{fig:mse_curve}
    \vspace{-4mm}
\end{figure}

The results in \autoref{fig:mse_curve} show that all previously discussed implementations exhibit very similar performance. For low \gls{snr} regions, the time-domain signal models also outperform multiharmonic signal model based estimators.

\section{Conclusion}
Application of the \gls{mle} for short periodic pulses has been computationally demanding/exhaustive due to the fine grid resolutions required for accurate results. 
We showed that by incorporating knowledge of the signal model, the number of grid points $G$ decouples from the number of measurements $N$ w.r.t. the runtime and memory complexity.
The "correlation and gather" approach also leverages recent optimizations in GPU architectures and drivers, allowing for strong efficiency gains.
The suggested implementation is therefore highly suitable for real-time tasks in industrial processes.

\section*{Acknowledgment}
The financial support by the Austrian Federal Ministry of Economy, Energy and Tourism, the National Foundation for Research, Technology and Development and the Christian Doppler Research Association is gratefully acknowledged.
\bibliographystyle{IEEEtran}
\bibliography{references}
\end{document}

%% file: fig/pulses.tex
\begin{tikzpicture}[
    >=latex, 
    font=\footnotesize,
    % Increase base line width to 1pt (slightly more than 'thick')
    axis/.style={->, >=latex, line width=1pt, darkgray},
    tick/.style={line width=1pt, darkgray},
    % 'pulse' is now 1.2pt to stand out
    pulse/.style={line width=1.2pt, samples=100, smooth},
    % Dashed lines disappear easily; set to 0.6pt minimum
    guide_line/.style={dash pattern=on 3pt off 2pt, draw=gray, line width=0.6pt},        
    dim_black/.style={black, <->, >=latex, line width=0.8pt}
]

    \pgfmathsetmacro{\A}{1.5}         
    \pgfmathsetmacro{\T}{2.8}         
    \pgfmathsetmacro{\Tp}{0.9}        
    \pgfmathsetmacro{\tauZero}{0.7}   
    \pgfmathsetmacro{\guideH}{2}    

    \draw[axis] (-0.3, 0) -- (8.0, 0) node[below left] {$t$};
    \draw[axis] (0, -0.3) -- (0, \guideH+0.15);
    \node[below left] at (0,0) {$0$};

    \pgfmathsetmacro{\nomZero}{\tauZero}

    \draw[tick] (\nomZero, 0.1) -- (\nomZero, -0.1) node[below, yshift=-1.5pt] {$\tau_0$};
    \draw[guide_line] (\nomZero, 0) -- (\nomZero, \guideH); 

    \draw[dim_black] (0, 0.4) -- (\nomZero, 0.4) node[midway, above=-1pt] {$\tau_0$};

    % Changed 'very thick' to explicit 1.5pt for the main signal
    \draw[pulse, green!60!black, line width=1.5pt] 
        plot[domain=0:\Tp] ({\x + \nomZero}, {\A * 0.5 * (1 - cos(360 * \x / \Tp))});

    \draw[dim_black] ({\nomZero + \Tp/2}, 0) -- ({\nomZero + \Tp/2}, \A) node[midway, fill=white, inner sep=1pt] {$A$};
    \draw[guide_line] ({\nomZero + \Tp}, 0) -- ({\nomZero + \Tp}, \guideH);
    \draw[dim_black] (\nomZero, \guideH-0.1) -- ({\nomZero + \Tp}, \guideH-0.1) node[midway, fill=white, inner sep=2pt] {$T_\text{p}$};
    \node[green!50!black, anchor=west] at ({\nomZero + \Tp}, \A*0.7) {$s_\text{p}(t)$};

    \pgfmathsetmacro{\nomOne}{\tauZero + \T}

    \draw[tick] (\nomOne, 0.1) -- (\nomOne, -0.1) node[below, yshift=0pt] {$T+\tau_0$};
    \draw[guide_line] (\nomOne, 0) -- (\nomOne, \guideH);

    \draw[pulse, black] 
        plot[domain=0:\Tp] ({\x + \nomOne}, {\A * 0.5 * (1 - cos(360 * \x / \Tp))});

    \pgfmathsetmacro{\nomTwo}{\tauZero + 2*\T}

    \draw[tick] (\nomTwo, 0.1) -- (\nomTwo, -0.1) node[below, yshift=0pt] {$2T+\tau_0$};
    \draw[guide_line] (\nomTwo, 0) -- (\nomTwo, \guideH);

    \draw[pulse, black] 
        plot[domain=0:\Tp] ({\x + \nomTwo}, {\A * 0.5 * (1 - cos(360 * \x / \Tp))});

    \draw[dim_black] (\nomOne, \guideH-0.1) -- (\nomTwo, \guideH-0.1) node[midway, fill=white, inner sep=2pt] {$T$};

    \node at (0,  \guideH+0.3) {$s(t;\boldsymbol{\theta})$};

\end{tikzpicture}

%% file: fig/drift.tex
\begin{tikzpicture}[
    >=latex, 
    font=\footnotesize,
    axis/.style={->, >=latex, line width=1pt, darkgray},
    tick/.style={line width=1pt, darkgray},
    pulse/.style={line width=1.2pt, green!60!black, samples=100, smooth},
    pulsedelta/.style={dash pattern=on 3pt off 2pt, line width=1.2pt, samples=100, smooth},
    guide_line/.style={dash pattern=on 3pt off 2pt, gray, line width=0.6pt},        
    dim_black/.style={black, <->, >=latex, line width=0.8pt},
    overlap/.style={fill=green!20}
]

    \pgfmathsetmacro{\A}{0.8}         % Reduced pulse amplitude
    \pgfmathsetmacro{\T}{2.8}
    \pgfmathsetmacro{\Td}{\T + 0.15}     
    \pgfmathsetmacro{\Tp}{1}        
    \pgfmathsetmacro{\tauZero}{0.7}   
    \pgfmathsetmacro{\guideH}{1.2}    % Reduced vertical guide line height

    % Calculate center point for the axis break
    \pgfmathsetmacro{\nomTwoD}{\tauZero + 2*\Td}
    \pgfmathsetmacro{\nomK}{\tauZero + 3*\T}
    \pgfmathsetmacro{\breakX}{(\nomTwoD + \Tp + \nomK) / 2}

    % Broken Axis with Spring/Zigzag
    \draw[line width=1pt, darkgray] (-0.3, 0) -- (\breakX - 0.25, 0);
    \draw[line width=1pt, darkgray, decorate, decoration={zigzag, segment length=2.5mm, amplitude=1.2mm}] (\breakX - 0.25, 0) -- (\breakX + 0.25, 0);
    \draw[axis] (\breakX + 0.25, 0) -- (11.75, 0) node[below left] {$t$};
    
    \node at (0, \guideH+0.3) {$s(t;\boldsymbol{\theta})$};
    \draw[axis] (0, -0.3) -- (0, \guideH+0.15);
    \node[below left] at (0,0) {$0$};

    \pgfmathsetmacro{\nomZero}{\tauZero}
    \pgfmathsetmacro{\nomOne}{\tauZero + \T}
    \pgfmathsetmacro{\nomTwo}{\tauZero + 2*\T}
    \pgfmathsetmacro{\nomZeroD}{\tauZero}
    \pgfmathsetmacro{\nomOneD}{\tauZero + \Td}
    \pgfmathsetmacro{\nomKD}{\nomK + \Tp + 0.2}

    % --- Pulse 0 ---
    \draw[tick] (\nomZero, 0.1) -- (\nomZero, -0.1) node[below, yshift=-1.5pt] {$\tau_0$};
    \draw[guide_line] (\nomZero, 0) -- (\nomZero, \guideH);
    
    \begin{scope}
    \clip plot[domain=0:\Tp, samples=100, smooth] ({\x + \nomZero}, {\A * 0.5 * (1 - cos(360 * \x / \Tp))}) -- ({\nomZero+\Tp},0) -- (\nomZero,0) -- cycle;
    \fill[overlap] plot[domain=0:\Tp, samples=100, smooth] ({\x + \nomZeroD}, {\A * 0.5 * (1 - cos(360 * \x / \Tp))}) -- ({\nomZeroD+\Tp},0) -- (\nomZeroD,0) -- cycle;
    \end{scope}
    \draw[pulse] plot[domain=0:\Tp] ({\x + \nomZero}, {\A * 0.5 * (1 - cos(360 * \x / \Tp))});
    \draw[pulsedelta] plot[domain=0:\Tp] ({\x + \nomZeroD}, {\A * 0.5 * (1 - cos(360 * \x / \Tp))});

    % --- Pulse 1 ---
    \begin{scope}
    \clip plot[domain=0:\Tp, samples=100, smooth] ({\x + \nomOne}, {\A * 0.5 * (1 - cos(360 * \x / \Tp))}) -- ({\nomOne+\Tp},0) -- (\nomOne,0) -- cycle;
    \fill[overlap] plot[domain=0:\Tp, samples=100, smooth] ({\x + \nomOneD}, {\A * 0.5 * (1 - cos(360 * \x / \Tp))}) -- ({\nomOneD+\Tp},0) -- (\nomOneD,0) -- cycle;
    \end{scope}
    \draw[tick] (\nomOne, 0.1) -- (\nomOne, -0.1) node[below, yshift=0pt] {$T+\tau_0$};
    \draw[guide_line] (\nomOne, 0) -- (\nomOne, \guideH);
    \draw[pulse]  plot[domain=0:\Tp] ({\x + \nomOne}, {\A * 0.5 * (1 - cos(360 * \x / \Tp))});
    \draw[guide_line] (\nomOneD, 0) -- (\nomOneD, \guideH);
    \draw[pulsedelta]  plot[domain=0:\Tp] ({\x + \nomOneD}, {\A * 0.5 * (1 - cos(360 * \x / \Tp))});
    
    % Outside-in dimension arrows for \delta (tightened vertically)
    \draw[black, line width=0.6pt] (\nomOne, \A+0.1) -- (\nomOne, \A+0.3);
    \draw[black, line width=0.6pt] (\nomOneD, \A+0.1) -- (\nomOneD, \A+0.3);
    \draw[black, <-, >=latex, line width=0.8pt] (\nomOne, \A+0.2) -- +(-0.35,0);
    \draw[black, <-, >=latex, line width=0.8pt] (\nomOneD, \A+0.2) -- +(0.35,0);
    \node[above, yshift=1pt] at ({(\nomOne+\nomOneD)/2}, \A+0.2) {$\delta$};

    % --- Pulse 2 ---
    \begin{scope}
    \clip plot[domain=0:\Tp, samples=100, smooth] ({\x + \nomTwo}, {\A * 0.5 * (1 - cos(360 * \x / \Tp))}) -- ({\nomTwo+\Tp},0) -- (\nomTwo,0) -- cycle;
    \fill[overlap] plot[domain=0:\Tp, samples=100, smooth] ({\x + \nomTwoD}, {\A * 0.5 * (1 - cos(360 * \x / \Tp))}) -- ({\nomTwoD+\Tp},0) -- (\nomTwoD,0) -- cycle;
    \end{scope}
    \draw[tick] (\nomTwo, 0.1) -- (\nomTwo, -0.1) node[below, yshift=0pt] {$2T+\tau_0$};
    \draw[guide_line] (\nomTwo, 0) -- (\nomTwo, \guideH);
    \draw[pulse] plot[domain=0:\Tp] ({\x + \nomTwo}, {\A * 0.5 * (1 - cos(360 * \x / \Tp))});
    \draw[guide_line] (\nomTwoD, 0) -- (\nomTwoD, \guideH);
    \draw[pulsedelta]  plot[domain=0:\Tp] ({\x + \nomTwoD}, {\A * 0.5 * (1 - cos(360 * \x / \Tp))});
    
    % Outside-in dimension arrows for 2\delta (tightened vertically)
    \draw[black, line width=0.6pt] (\nomTwo, \A+0.1) -- (\nomTwo, \A+0.3);
    \draw[black, line width=0.6pt] (\nomTwoD, \A+0.1) -- (\nomTwoD, \A+0.3);
    \draw[black, <-, >=latex, line width=0.8pt] (\nomTwo, \A+0.2) -- +(-0.35,0);
    \draw[black, <-, >=latex, line width=0.8pt] (\nomTwoD, \A+0.2) -- +(0.35,0);
    \node[above, yshift=1pt] at ({(\nomTwo+\nomTwoD)/2}, \A+0.2) {$2\delta$};

    % Break Indicator (Ellipsis)
    \node[darkgray] at (\breakX, \A*0.5) {\Large $\dots$};

    % --- Pulse k ---
    \begin{scope}
    \clip plot[domain=0:\Tp, samples=100, smooth] ({\x + \nomK}, {\A * 0.5 * (1 - cos(360 * \x / \Tp))}) -- ({\nomK+\Tp},0) -- (\nomK,0) -- cycle;
    \fill[overlap] plot[domain=0:\Tp, samples=100, smooth] ({\x + \nomKD}, {\A * 0.5 * (1 - cos(360 * \x / \Tp))}) -- ({\nomKD+\Tp},0) -- (\nomKD,0) -- cycle;
    \end{scope}
    \draw[tick] (\nomK, 0.1) -- (\nomK, -0.1) node[below, yshift=0pt] {$kT+\tau_0$};
    \draw[guide_line] (\nomK, 0) -- (\nomK, \guideH);
    \draw[pulse] plot[domain=0:\Tp] ({\x + \nomK}, {\A * 0.5 * (1 - cos(360 * \x / \Tp))});
    \draw[guide_line] (\nomKD, 0) -- (\nomKD, \guideH);
    \draw[pulsedelta]  plot[domain=0:\Tp] ({\x + \nomKD}, {\A * 0.5 * (1 - cos(360 * \x / \Tp))});
    
    % Standard dimension arrows for larger k\delta gap (tightened vertically)
    \draw[dim_black] (\nomK, \A+0.2) -- ({\nomKD }, \A+0.2) node[midway, above, inner sep=2pt] {$k\delta$};

\end{tikzpicture}